\documentclass[prl,superscriptaddress,amsmath,amssymb,showpacs,noshowkeys,a4paper,twocolumn]{revtex4}

\usepackage{graphicx}
\usepackage{dcolumn}
\usepackage{bm}
\usepackage[usenames]{color}
\usepackage{epstopdf}
\definecolor{mygray}{gray}{0.5}

\newcommand{\affmsc}{\affiliation{Mati\`ere et Syst\`emes Complexes, Universit\'e Paris Diderot, CNRS - UMR 7057, B\^atiment Condorcet, 10 rue Alice Domon et L\'eonie Duquet, 75013 Paris, France}}

\newcommand{\affmpq}{\affiliation{Institut Langevin, ESPCI ParisTech and Universit\'e Paris Diderot, CNRS UMR 7587, 10 rue Vauquelin, 75231 Paris Cedex 05, France}}

\newcommand{\afffast}{\affiliation{Laboratoire de Physique Statistique,
Ecole Normale Sup\'erieure and Universit\'e Pierre et Marie Curie, 24 rue Lhomond, 75231 Paris Cedex 05, France}}

\begin{document}

\title{Mutual adaptation of a Faraday instability pattern with its flexible boundaries in floating fluid drops\footnote{G.~Pucci, E.~Fort, M.~Ben Amar and Y.~Couder,
{\em Phys. Rev. Lett.}, \textbf{106} 024503, (2011).\\
http://link.aps.org/doi/10.1103/PhysRevLett.106.024503. \\ Copyright of American Physical Society.}}

\author{G.~Pucci}
\affmsc

\author{E.~Fort}
\affmpq

\author{M.~Ben Amar}
\afffast

\author{Y.~Couder}
\affmsc

\begin{abstract}
Hydrodynamic instabilities are usually investigated in confined geometries where the resulting spatio-temporal pattern is constrained by the boundary conditions. Here we study the Faraday instability in domains with flexible boundaries. This is implemented by triggering this instability in a floating fluid drop. An interaction of  Faraday waves with the shape of the drop is observed, the radiation stress of the waves exerting a force on the surface tension held boundaries. Two regimes are observed. In the first one there is a co-adaptation of the wave pattern with the shape of the domain so that a steady configuration is reached. In the second one the radiation stress dominates and no steady regime is reached. The drop stretches and ultimately breaks into smaller domains that have a complex dynamics including spontaneous propagation.
\end{abstract}
\pacs{05 65.+b,  Self-organized systems,  05 45.-a, Non linear dynamics and chaos, 47.20.Ma
Interfacial instabilities , 47.35.Pq, Capillary waves}

\maketitle

\begin{figure}[hbtp]
\includegraphics[width=.68\linewidth]{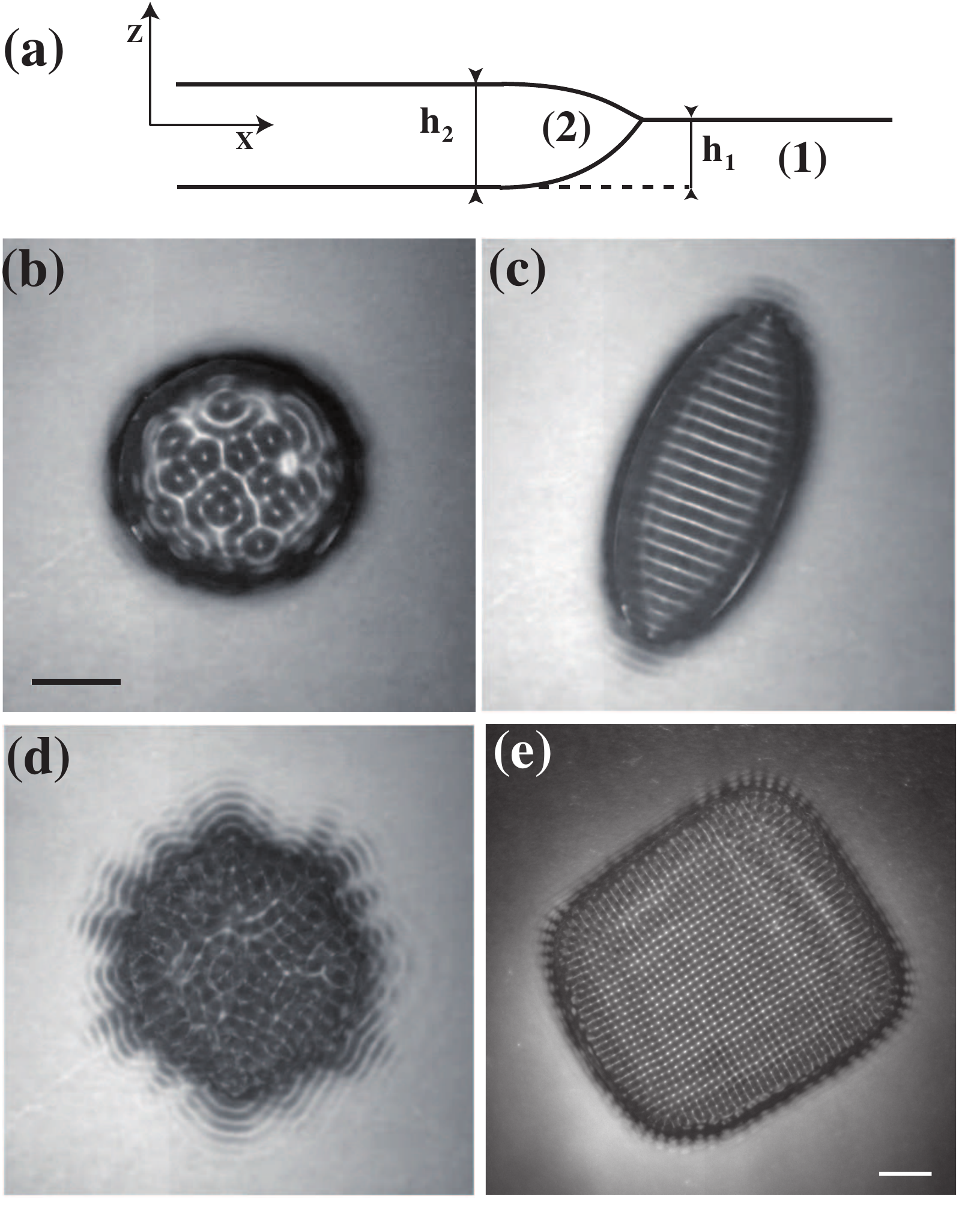}
\caption{(a) Vertical section of an isopropanol drop  floating on perfluorated oil  in absence of oscillations. (b-d) A drop of volume $V_2 = 1.00 \pm 0.02 \mathrm{ml}$ forced at $f_0 = 130\mathrm{Hz}$ for three amplitudes of forcing. (b) Faraday waves of weak amplitude, $\gamma_{m}/g  = 3.28$, (c) steady elongated state, $\gamma_{m}/g =3.93$ and (d) highly perturbed state $\gamma_{m}/g = 6.48$. (e) A drop of volume $V_2 = 10.0 \pm 0.1 \mathrm{ml}$ square when forced at $f_0= 160\mathrm{Hz}$ with $\gamma_{m}/g =4.90$. The bars are 1cm long.}
\label{fig:1}
\end{figure}

Fluid dynamics instabilities usually appear in two types of situations corresponding to confined or opened geometries respectively. For instance thermal buoyancy, when confined in a box, gives rise to patterns of Rayleigh-B\'enard rolls adapted to the boundaries \cite{Getling}. 
In contrast in an open medium, an isolated source of heat generates a thermal plume in which the growing turbulent structures define the envelop of the unstable region \cite{Zocchi, BenAmar}. 
We address an intermediate situation in which an instability develops inside a finite domain with flexible boundaries and we study the interplay between the pattern and its borders. We use the Faraday instability in which waves form on the surface of a fluid submitted to vertical oscillations \cite{Faraday} and whose resulting patterns have been widely studied in confined geometries \cite{Douady, Edwards}. Here we choose to confine the Faraday instability in a drop of low viscosity floating on a very viscous, stable, immiscible fluid [Fig.\ref{fig:1}(a)].
We use a classical Faraday experiment set-up. A circular cell (radius 10cm, depth 0.8cm) is placed on a vibration exciter generating a vertical oscillation of acceleration $\gamma(t) = \gamma_m \cos(2\pi f_0 t)$. The investigated frequency and amplitude ranges are respectively $50\mathrm{Hz} < f_0 < 250\mathrm{Hz}$ and $0 < \gamma_m /g < 10$, where g is the acceleration of gravity. The motion can be observed by a stroboscopic video camera or a high speed video camera.
The heavier and more viscous fluid (fluid 1) fills the cell and forms a bath of thickness 5mm.  A controlled quantity $V_2$ of the less viscous fluid (fluid 2) is then deposited becoming a floating drop that  forms a single circular pancake far from the boundaries. Its radius  $r_0$ and thickness $h_2$ at rest result from the equilibrium between buoyancy and capillarity \cite{Langmuir, Noblin}. The size of the bordering  meniscus fixed by the capillary length is small compared to the horizontal size. In all our experiments we find $h_2\simeq2$mm and $r_0\simeq1.3$cm.

\begin{figure}[hbtp]
\includegraphics[width=0.68\linewidth]{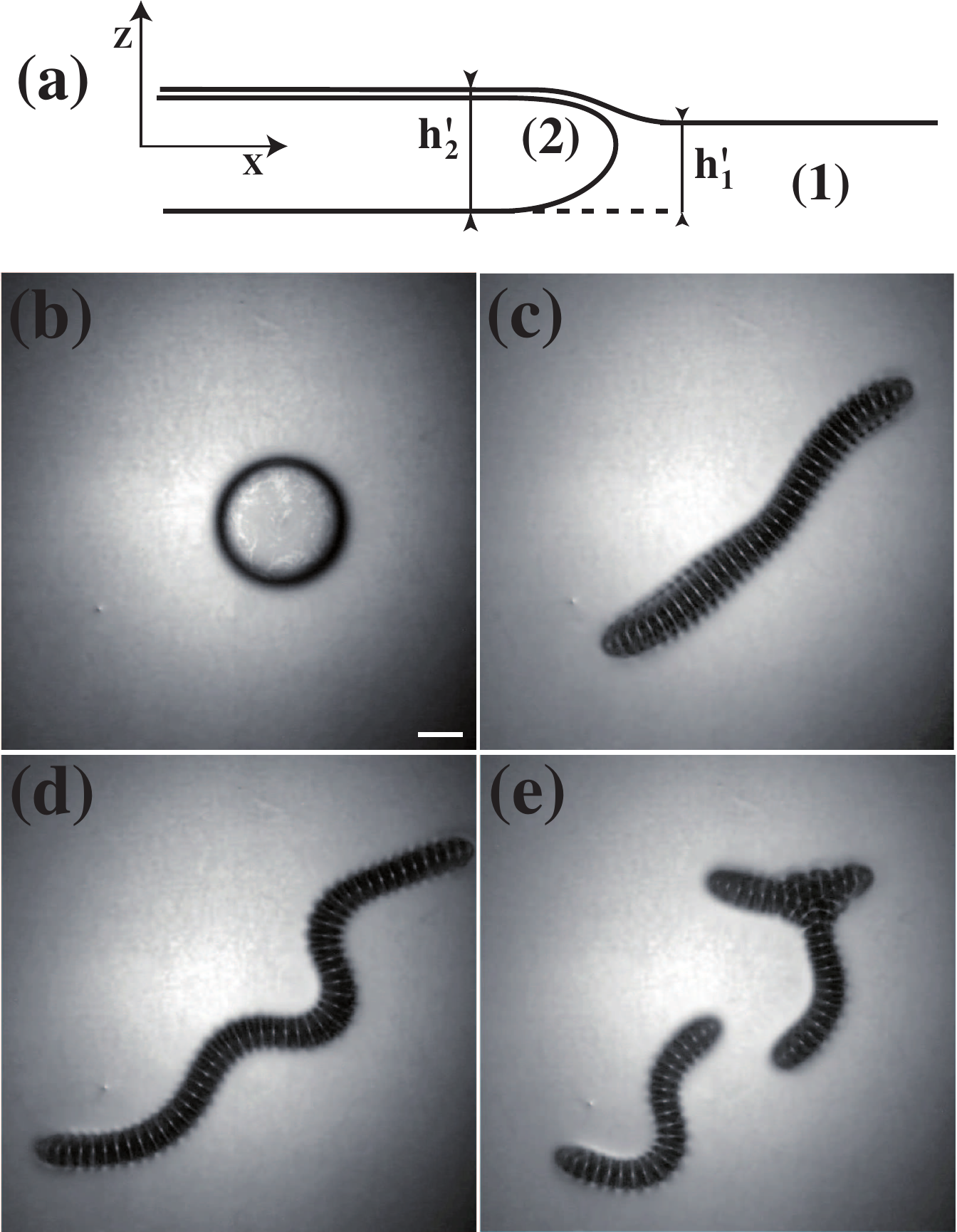}
\caption{The case of an ethanol drop of volume $V^{\prime}_2 = 1.00 \pm 0.02$ml deposited on silicon oil.  (a) Sketch of the vertical section of the floating drop in the absence of oscillations. (b) It is circular at rest with area and thickness $A_2 = 459 \pm 5 \mathrm{mm^2}$ and  $h_2 = 2.2$mm respectively. (c-e) Three successive images showing the temporal evolution of the drop when forced at $f_0 = 130\mathrm{Hz}$ with $\gamma_m /g = 7.00$. The bar is 1cm long. }
\label{fig:2}
\end{figure}

There is a range of forcing amplitude for which the Faraday instability forms in the drop only. The resulting waves are initially disordered and generate fluctuations of the drop boundary [Fig.\ref{fig:1}(b)]. In return the fluctuations of the boundary generate an unsteadiness of the wave pattern. Two possible archetypes of evolution can then be observed [Fig.\ref{fig:1}-\ref{fig:2}] that depend on the fluids. We investigated many pairs of immiscible fluids, having different viscosity contrasts and wetting properties, and found only these two possible behaviours.

In the first archetype, the system self-organizes by a co-evolution of the wave field and the boundaries so that an equilbrium is reached \cite{movie1}. The example shown in fig.\ref{fig:1} was obtained with isopropanol (density $\rho_2 = 785 \mathrm{kg/m^3}$, viscosity $\mu_2 = 1.8 \times 10^{-3} \mathrm{Pa\cdot s}$ and surface tension against vapor $\sigma_2 = 24 \mathrm{mN/m}$) floating on perfluorated oil ($\rho_1 = 1850 \mathrm{kg/m^3}$, $\mu_1 = 26 \times 10^{-3} \mathrm{Pa \cdot s}$ and  $\sigma_1 = 15 \mathrm{mN/m}$). The measured interfacial tension is $\sigma_{1,2} = 6.3 \mathrm{mN/m}$. 

In the second type of evolution   the displacement of the boundary due to the waves does not lead to equilibrium \cite{movie2}. The formation of parallel standing waves results into a constantly increasing elongation of the drop into a snake-like structure which breaks into fragments having a large variety of dynamical behaviors. The example shown in fig.\ref{fig:2} was obtained with ethanol ($\rho^{\prime}_2 = 789 \mathrm{kg/m^3}$, $\mu^{\prime}_2 = 0.9 \times 10^{-3} \mathrm{Pa \cdot s}$, $\sigma^{\prime}_2 = 23 \mathrm{mN/m}$) floating on silicon oil ($\rho^{\prime}_1 = 965 \mathrm{kg/m^3}$, $\mu^{\prime}_1 = 100 \times 10^{-3} \mathrm{Pa \cdot s}$ and  $\sigma^{\prime}_1 = 20 \mathrm{mN/m}$). The measured interfacial tension is $\sigma^{\prime}_{1,2} = 0.7\mathrm{mN/m}$.

The difference between the two regimes can be understood by a dimensional analysis evaluating the ratio of the destabilizing factor (the radiation pressure) to the restoring one (capillary pressure) as will be discussed below.

\begin{figure}[hbtp]
\includegraphics[width=.7\linewidth]{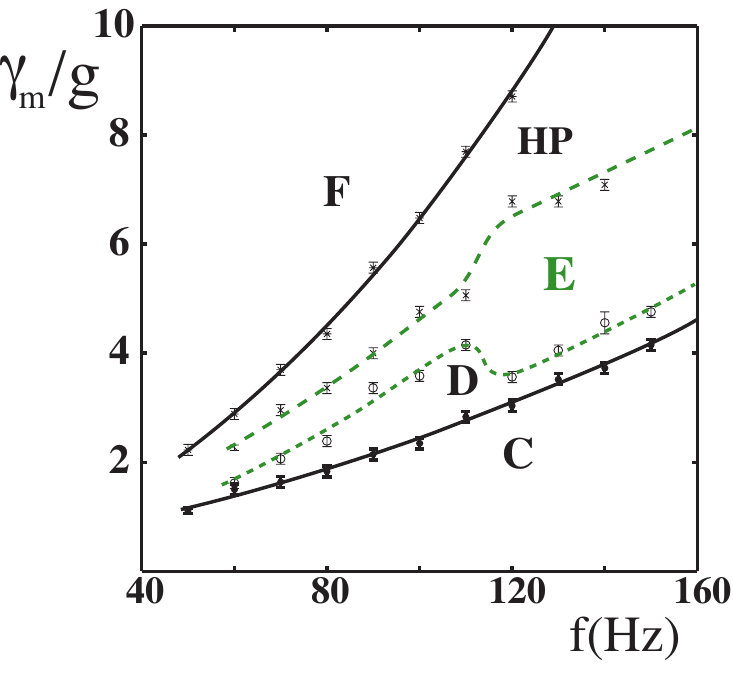}
\caption{Phase diagram of the different regimes observed in the case of an isopropanol drop of volume $V_2 = 1.00 \pm 0.02$ ml deposited on perfluorated oil. The drop area at rest is $A_2 = 531 \pm 11 \mathrm{mm^2}$, its average thickness $h_2 = 1.9$mm. C = circular, D = deformed, E = elongated, HP = highly perturbed, F = Faraday instability in oil.}   
\label{fig:3}
\end{figure}

\begin{figure}[hbtp]
\includegraphics[width= .62\linewidth]{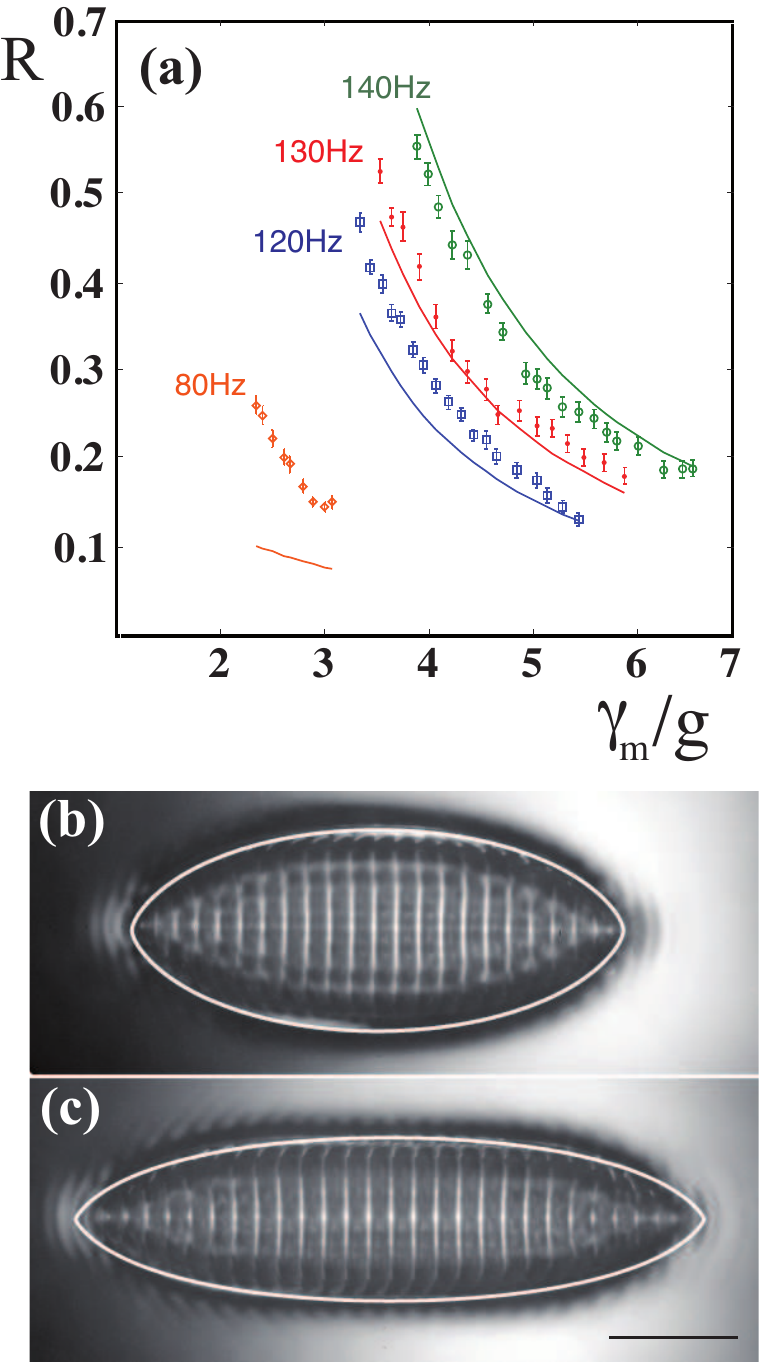}
\caption{(a) Evolution of the aspect ratio of an elongated isopropanol drop on perfluorated oil as a function of the forcing amplitude for different frequencies. The lines are the fits by eq.\ref{eqn:4}. (b, c) Photographs of two drops in an elongated state at $f_0 = 130$Hz for $\gamma_m / g = 3.97$ and $R = 0.42$ (b) and $\gamma_m / g = 4.62$ and $R = 0.28$ (c) and the predicted shapes by the corresponding solutions [Eq.\ref{fig:3}]. The bar is 1cm long.}   
\label{fig:4}
\end{figure}

We first characterize experimentally  the former archetype as a function of the frequency and the forcing amplitude. 
A phase diagram depicting the different states of the system is shown in fig.\ref{fig:3}. When the forcing amplitude is increased, Faraday waves appear in the drop above a first onset ($\gamma_m > \gamma^{D}_m$). They form complex unsteady patterns which deform the drop [Fig.\ref{fig:1}(b)], its average shape remaining circular. 
Above a second threshold $\gamma^{E}_m$ the waves strengthen and start organizing, resulting into an elongation of the drop in the perpendicular direction. Correlatively this elongation favors a further organization of the waves. By a slow evolution the drop reaches a stable elongated shape and the waves become steady too [Fig.\ref{fig:1}(c)]. They are perpendicular to the long sides of the drop (as usual for Faraday waves in elongated cells). By increasing the forcing the drop undergoes a further elongation to a new steady state. The shape can be defined by the ratio $R$ of the length of the minor over the major axis of the drop. Fig.\ref{fig:4}(a) shows the evolution of $R$ as a function of $\gamma_m$ for various frequencies. Above a third threshold $\gamma^{HP}_m$, the waves become chaotic and the envelope of the domain undergoes very large fluctuations [Fig.\ref{fig:1}(d)]. Finally at a fourth threshold $\gamma^{F}_m$ Faraday waves form on the viscous substrate. We cannot obtain the second archetype by increasing the forcing amplitude. We measured the area $A_2$ covered by the drop during this process and found that it does not change. The observed wavelength is given by the dispersion relation of gravity-capillary waves for the Faraday frequency $f_0 /2$. 

The first type of behavior can be understood as resulting from the effect of the radiation pressure $P_r$ \cite{LonguetHiggins} of the surface waves distorting the boundaries of the pancake. We thus seek stable solutions in which the radiation and hydrostatic pressures are balanced by capillarity. We assume an unidirectional standing wave along the x-axis, with small thickness for the drop and small wavelength for the wave, the initial radius of the drop being the length unit. The small thickness assumption allows a lubrication approximation transforming the 3D boundary value problem in a 2D one while the small wavelength approximation allows an average on larger length-scales. An asymptotic analysis adapted from \cite{Burgess} shows that the function $y(x)$ describing the drop shape satisfies a two-dimensional Laplace law modified by the radiation pressure along the normal. It is a Riccati equation:

\begin{equation}
\label{eqn:1}
P_h + P_{r} \frac{y^{\prime 2}(x)}{1+y^{\prime 2}(x)} = - \sigma_2 f(\theta) \frac{d}{dx} \frac{y^{\prime} (x)}{\sqrt{1+y^{\prime 2 }(x)}}
\end{equation}
where $P_h$ is a term of hydrostatic pressure (assumed constant for quasi-steady boundary), $f(\theta)$ depends on the wetting angle and $P_r$ is the radiation pressure
\begin{equation}
\label{eqn:2}
P_r = \frac{\rho_2}{8} \frac{\rho_1 - \rho_2}{\rho_1} \omega^2 A^2 F^2
\end{equation}
with $A$ the amplitude of the Faraday waves,  $\omega = \pi f_0$ the angular frequency and $F = e^{h_1-h_2} (1 + B \cosh kh_2) - B \sinh k h_1$, with $B= - k/{\omega^2} [(\rho_2 - \rho_1)/{\rho_2} g -\sigma_{1,2} k^2 /{\rho_1}]$, a factor that takes into account thickness effects and waves transmission for a baroclinic mode in fluid 1, $k$ being the wave vector modulus. Using the volume conservation, the parameters of eq.\ref{eqn:1} reduce only to one free parameter responsible for the final equilibrium shape, that is the ratio $a = P_r / P_h$. In this case we have found explicit analytical solutions of eq.\ref{eqn:1}. The drop shape is given by  
\begin{equation}
\label{eqn:3}
\begin{split}
y(x)  =  \pm \frac{1}{P_{h}' \sqrt{1+a}} \log [ \sqrt{1+a}( \cos P_{h}' \sqrt{a} x \:+ \\
  + \: \sqrt{\sin^2 \Phi_0 - \sin^2 P_{h}' \sqrt{a} x}) ] 
\end{split}
\end{equation}
with $ \sin^2 \Phi_0 = a/(1+a) $ and $ P_{h}' = P_h /[\sigma_2 f(\theta)] $. The conservation of the area gives $P_{h}' = [ \log(1+a) / a \sqrt{1+a} ]^{1/2}$, being $P_{h}' = 1$ at rest ($P_r = 0$).
When $R>0.25$ good fits of the experimental shapes are obtained [Fig.\ref{fig:4}(b, c)]. The evolution of the aspect ratio $R$ with forcing can also be obtained analytically
\begin{equation}
\label{eqn:4}
R = \frac{\log \left( \sqrt{a} + \sqrt{1+a}\right)}{\sqrt{1+a} \arcsin \sqrt{a/(1+a)}}
\end{equation}
We measured the amplitude $A$ of the waves in the drops by lateral zoomed-in films and checked that it varies as $(\gamma_m - {\gamma^{D}_m})^{1/2}$. Using it  to evaluate $P_r$ and thus $a$, we obtain fits of $R(\gamma_m)$ for various frequencies [Fig.\ref{fig:4}(a)].
The computed shapes become inconsistent with the initial small wavelength and small thickness hypothesis of the model whenever the radius of curvature at the tip of the computed shape becomes small. For this reason the very elongated shapes ($R<0.25$) obtained with this model cannot account for reality. Experimentally we observe that the shape of the tip becomes fixed while the drop keeps elongating, the curvature radius of the tip being of the order of the wavelength. 

We can now turn to the second archetype. The main characteristic is that ethanol is wetted by silicon oil. At rest an oil film is observed to cover the upper surface of the drop [Fig.\ref{fig:2}(a)]. The instability in ethanol appears in a subcritical way i.e. large amplitude waves form at threshold. These waves stretch the drop [Fig.\ref{fig:2}(c)] but there is no convergence towards a final stable shape. There is no theory for this dynamical regime yet.

However its existence can be understood by dimensional analysis. We use $a_0= P_r/ P_{h}^{0} $, the ratio of the estimated wave radiation pressure [Eq.\ref{eqn:2}] to the 2D pressure due to capillary effects in the circular drop at rest. For drops of centimetric size and Faraday waves of the usually observed amplitude this analysis applied to the first archetype gives $a_0 \simeq 0.1$. In the second archetype, because of the wetting, the capillary tension is the interfacial one $\sigma^{\prime}_{1,2} = 0.7\mathrm{mN/m}$. As a result we find $a_{0}' \simeq 2$. While in the first case the pressure exerted by the waves is only a perturbation of the capillary equilibrium of the drop, in the second this radiation pressure exceeds the possible response of capillary forces so that no steady solution can be reached.  

\begin{figure}[hbtp]
\includegraphics[width=.65\linewidth]{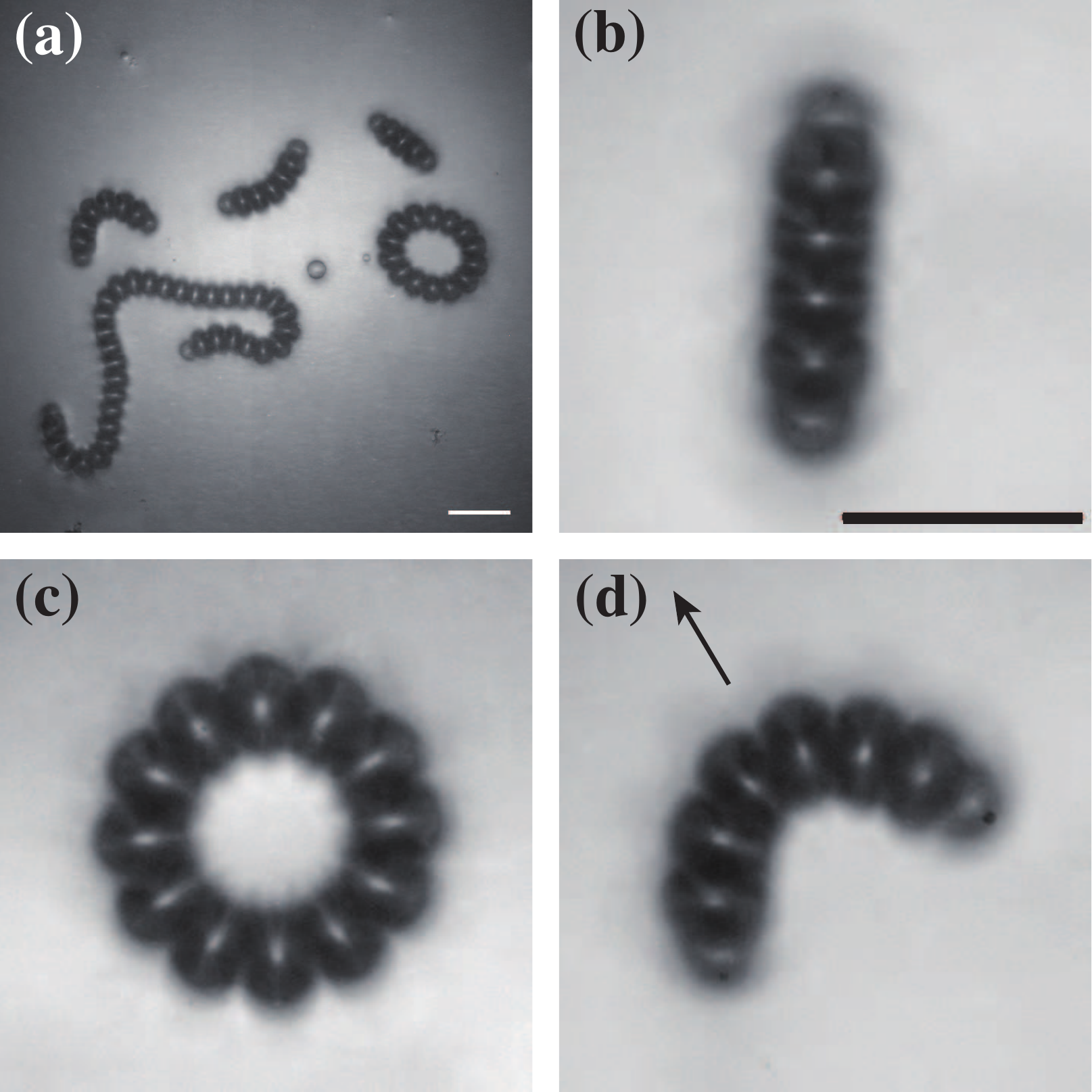}
\caption{(a) The surface of the experimental cell after the fragmentation of a drop of ethanol on silicon oil for $f_0 = 80$Hz and $\gamma_m /g = 5.06$. (b-d) The shape of some of the resulting steady fragments.  Each one corresponds to a different fragment size. The bars are 1cm long. }
\label{fig:5}
\end{figure}

The elongation is followed by buckling [Fig.\ref{fig:2}(d)] then breaking into several fragments [Fig.\ref{fig:2}(e), Fig.\ref{fig:5}(a)]. We measured the total area $A^{\prime}_2$ covered by the pattern and found that it increases with forcing, proving that the drops are stretched by the waves (their thickness decreases). By buckling  large  fragments can  take  croissant or a horseshoe shapes [Fig.\ref{fig:5}(d)]. These shapes are stationary but propagate in the direction of the curvature. This is one more example of self-propagation by a spontaneous symmetry breaking \cite{Protiere}. The velocity is constant and a croissant (or a horseshoe) could move indefinitely in an infinite bath. In practice a bath of finite size is covered with moving and motionless fragments [Fig.\ref{fig:5}(a)] that keep colliding, merging and splitting. The global aspect is reminiscent of the interplay of structures obtained in cellular automata such as Conway's game of life \cite{Gardner}. Very small fragments are observed to remain steady with a stationary shape that recalls the elongated one of the first regime [Fig.\ref{fig:5}(b)]. This possibility of a steady regime for small drops can be understood by the dimensional analysis : $ a_0$ becomes smaller when the radius of the drop at rest is very small.  

A remarkable feature of the snake-like structure is that they have constant transverse widths and tip radii, both being of the order of the Faraday wavelength. Their stretching and instabilities affect mostly the middle of their length. The buckling could be due to the generation of a streaming flow \cite{LonguetHiggins2} that has been already observed in Faraday instability \cite{Higuera}. This streaming effect could also be responsible for their motion on the substrate. The investigation of the resulting dynamical regimes is beyond the scope of this letter. 

We have found that, by slow dynamics, a mutual adaptation is possible between an instability and its boundaries. This phenomenon is related to the self-tuning of oscillators. In optical cavities the radiation pressure was shown \cite{Meystre} to create a coupling between the mirrors degrees of freedom and the optical field. Similarly, a forced mechanical oscillator with an additional degree of freedom \cite{Boudaoud} exhibits a slow drift by which it can self-tune. The present phenomenon is more general and should show up in other type of instabilities confined in adaptable boundaries.

We are grateful to Antonin Eddi and Julien Sebilleau for useful discussions and to Mathieu Receveur and Laurent R\'ea for technical assistance. Universit\'e Franco-Italienne (UFI) and ANR Blanche 02 97/01 supported this work.

\end{document}